\documentclass[]{aastex63}

\newcommand{\revision}{}
\newcommand{\secrev}{}
\newcommand{\newrev}{}

\received{May 25, 2021}
\accepted{February 9, 2022}
\submitjournal{PSJ}
\usepackage{subfigure}
\usepackage{graphicx}

\shorttitle{Detectability of Chlorofluorocarbons in the Atmospheres of Habitable M-dwarf Planets}
\shortauthors{Haqq-Misra et al.}
\graphicspath{{./}{figures/}}

\begin{document}

\title{Detectability of Chlorofluorocarbons in the Atmospheres of Habitable M-dwarf Planets}

\correspondingauthor{Jacob Haqq-Misra}
\email{jacob@bmsis.org}

\author{Jacob Haqq-Misra}
\affiliation{Blue Marble Space Institute of Science, Seattle, WA, USA}

\author{Ravi Kopparapu}
\affiliation{NASA Goddard Space Flight Center, 8800 Greenbelt Road, Greenbelt, MD 20771, USA}
\affiliation{Sellers Exoplanet Environment Collaboration (SEEC), NASA Goddard Space Flight Center}

\author{Thomas J. Fauchez}
\affiliation{NASA Goddard Space Flight Center, 8800 Greenbelt Road, Greenbelt, MD 20771, USA}
\affiliation{Goddard Earth Sciences Technology and Research (GESTAR), Universities Space Research Association, Columbia, MD, USA}
\affiliation{American University, Washington DC, USA}
\affiliation{Sellers Exoplanet Environment Collaboration (SEEC), NASA Goddard Space Flight Center}

\author{Adam Frank}
\affiliation{Department of Physics and Astronomy, University of Rochester, Rochester, New York, 14620}

\author{Jason T.\ Wright}
\affiliation{Department of Astronomy \& Astrophysics, The Pennsylvania State University, University Park, PA, 16802, USA}
\affiliation{Center for Exoplanets and Habitable Worlds, The Pennsylvania State University, University Park, PA, 16802, USA}
\affiliation{Penn State Extraterrestial Intelligence Center, The Pennsylvania State University, University Park, PA, 16802, USA}

\author{Manasvi Lingam}
\affiliation{Department of Aerospace, Physics and Space Sciences, Florida Institute of Technology, Melbourne, FL 32901, USA}

\begin{abstract}

The presence of chlorofluorocarbons (CFCs) in Earth's atmosphere is a direct result of technology. Ozone-depleting CFCs have been banned by most countries, but some CFCs have persistent in elevated concentrations due to their long stratospheric lifetimes. CFCs are effective greenhouse gases and could serve as a remotely detectable spectral signature of technology. Here we use a three-dimensional climate model and a synthetic spectrum generator to assess the detectability of CFC-11 and CFC-12 as a technosignature on exoplanets. We consider the case of TRAPPIST-1e as well as a habitable Earth-like planet around a 3300\,K M-dwarf star, with CFC abundances ranging from one to five times present-day levels. Assuming an optimistic James Webb Space Telescope (JWST) Mid Infrared Instrument (MIRI) low resolution spectrometer (LRS) noise floor level of 10\,ppm to multiple co-added observations, we find that spectral features potentially attributable to present or historic Earth-level CFC features could be detected with a SNR $\ge 3-5$ on TRAPPIST-1e, if present, in $\sim 100$ hours of in-transit time. However, applying a very conservative 50\,ppm noise floor to co-added observations, even a 5x Earth-level CFC would not be detectable no matter the observation time. Such observations could be carried out simultaneously and at no additional cost with searches for biosignature gases.  Non-detection would place upper limits on the CFC concentration. We find that with the launch of JWST, humanity may be approaching the cusp of being able to detect passive atmospheric technosignatures equal in strength to its own around the nearest stars.

\end{abstract}



\section{Introduction} \label{sec:intro}

Thousands of exoplanets have so far been discovered from space-based telescopes, such as Kepler and TESS, as well as ground-based observatories. Detection methods can constrain the orbital position and bulk properties of such planets, but follow-up observations of planetary spectra in transit, \secrev{emitted,} or reflected light can provide information about the presence and composition of a planet's atmosphere. These methods have already been demonstrated for the spectral characterization of gas giant atmospheres \citep[e.g.][]{dave2002,paper12019,paper22019}, while detecting and characterizing the atmospheres of smaller, Earth-sized planets remains an ongoing priority for exoplanet science. 

One of the astrobiological motivations for the spectral characterization of planetary atmospheres is the possibility of detecting evidence of life on an exoplanet. This has inspired the search for ``biosignatures,'' which refer to remotely detectable spectral features that could indicate evidence of life on an exoplanet. The idea of searching for spectral biosignatures has received significant attention with regard to identifying plausible biosignatures, assessing their detectability limits, and developing strategies for \newrev{conducting} \secrev{their} search  in tandem with the broader goals of the astrophysics community \citep[e.g.][]{seager2012astrophysical, kaltenegger2017characterize, grenfell2017review,Schwieterman2018, Meadows2018, Catling2018, Walker2018, Fujii2018, o2019expanding, lammer2019role}. 

Biosignatures refer generally to any remotely detectable evidence of life, while ``technosignatures'' \citep{tarter07} specifically describe observational evidence of technology that could be detected through astronomical means. Technosignatures are a logical continuation of the search for biosignatures, both of which draw upon the history of life and technology on Earth as examples of planetary evolution \citep{2018arXiv181208681N}. The science of identifying and classifying technosignatures, and developing cost-efficient methods to search for them, remains in a state of infancy compared to biosignature science \citep{wright2019,jacob2020,Lingam21}. Nevertheless, several possible classes of technosignatures have already been identified that include waste heat \citep{dyson60,carrigan09b,GHAT2,kuhn2015global}, artificial illumination \citep{Schneider2010,Loeb11,Kipping16,tabor2021detectability}, artificial atmospheric constituents \citep{owen1980search,2006dies.conf..247C,Schneider2010,Lin2014,Stevens16,kopparapu2021}, artificial surface constituents \citep{Lingam17}, stellar pollution \citep{shklovskii1966,Whitmire80,Stevens16}, non-terrestrial artifacts \citep{BRACEWELL1960,freitas1980search,Rose2004,JR2012}, and megastructures \citep{dyson60,Arnold05,Forgan13,GHAT4}. 

The modern era of SETI \citep[i.e., the search for technosignatures]{WrightAdHoc} began with the realization that humanity could use existing technology to detect a similar level of technology over interstellar distances \citep[i.e., powerful, deliberately directed radio signals]{SETI}. A major milestone in SETI would be achieved when \revision{present-day} detection technologies become sensitive enough to detect humanity’s ongoing and passive technosignatures at such distances. The full Square Kilometer Array, for instance, is thought to be sensitive enough to detect humanity’s typical radar emission at distances of a few parsecs \citep{Loeb2007}\revision{; however, future projections in which Earth's radio leakage decreases significantly would be much more difficult to detect \citep{forgan2011failure}.}

Industrial pollution represents a class of atmospheric constituents on Earth that could conceivably be technosignatures if observed in the spectra of an exoplanet. One example is nitrogen dioxide (NO$_2$), which has large sources on Earth from combustion that are greater than non-anthropogenic sources. A study by \citet{kopparapu2021} showed that the absorption features of NO$_2$ in the $0.2-0.7\,\mu$m range could be detectable with the Large Ultraviolet Optical Infrared Surveyor (LUVOIR, \citet{Fischer2019}). \citet{kopparapu2021} found that a 15\,m LUVOIR-like telescope could detect Earth-like levels of NO$_2$ for a planet around a Sun-like star with $\sim$400 hours of observation, while planets orbiting K-dwarf stars would require even less time due to the reduction in loss of NO$_2$ from photolysis in such systems. The detection of elevated NO$_2$ levels in the atmosphere of an exoplanet could be consistent with ongoing industrial processes on the surface, although any such observations would need to be evaluated against possible non-technological explanations before concluding that the NO$_2$ must be a technosignature. In this regard, such a search for technosignatures shares the same ambiguity or tentative nature as many or most searches for biosignatures.

In this study, we examine halocarbons---molecules that contain carbon and halogen atoms---as another class of pollutants that could serve as technosignatures. We specifically focus on chlorofluorocarbons (CFCs), which are only produced in significant quantities on Earth from industrial uses as refrigerants, blowing agents, and cleaning agents. CFCs are potent greenhouse agents with long atmospheric residence times. The only sink for most CFCs is photolysis by ultraviolet radiation in the stratosphere, which releases chlorine atoms that cause the depletion of stratospheric ozone on Earth \revision{\citep{seinfeld2008atmospheric}}. The Montreal Protocol of 1987 placed limits on the production of certain CFCs in order to prevent further damage to the ozone layer \citep{velders2007importance}. These provisions have been successful to the extent that ozone-depleting compounds are much less present in the stratosphere; however, the recovery of the ozone layer to pre-1980's levels appears to be slow and shows large uncertainties in both measurements and model projections \citep[e.g.][]{eyring2010multi,chipperfield2017detecting}. 

Observing CFCs in the atmosphere of an exoplanet would be compelling evidence of a technosignature. The accumulation of CFCs on an exoplanet could be the result of ongoing industrial processes \citep{owen1980search,2006dies.conf..247C,Schneider2010}, particularly for a planet on which ozone loss is not a concern. CFCs could also be useful in artificially increasing the greenhouse effect on a planet, which could have applications in terraforming a planet to increase its suitability for life \citep{marinova2005radiative,dicaire2013}. The detectability of CFCs for a planet orbiting a white dwarf host star was examined by \citet{Lin2014}, who focused specifically on CCl$_3$F (known as CFC-11) and CF$_4$ (known as CFC-14) because they both show strong absorption features in the infrared. \citet{Lin2014} estimated that CFCs at abundances ten times greater than present-day Earth could be detected in a white dwarf system by the James Webb Space Telescope (JWST) with $\sim 1.7$\,days of observing time. 

Here we calculate detectability limits of CFCs for an Earth-sized planet orbiting an M-dwarf star. Such systems are likely targets for characterization by upcoming space missions such as JWST or mission concepts such as LUVOIR, the Habitable Exoplanet (HabEx, \citet{Gaudi2019}), Origins (\citet{Cooray2019}), or Large Interferometer for Exoplanets (LIFE, \citet{Quanz2020life,2021arXiv210107500L}), in addition to large ground-based observatories like \secrev{Extremely} Large Telescopes (ELTs) \citep{snellen2015, quanz2015}. We focus on the detectability of CFC-11 \secrev{(CCl$_{3}$F)} and CFC-12 (CCl$_{2}$F$_{2}$), which are two of the most abundant CFCs in Earth's atmosphere with elevated levels that have persisted despite the Montreal Protocol. Our model simulations are constrained to planets within the habitable zone of the host star, which represents the circumstellar region where a terrestrial planet could sustain surface liquid water \citep{kasting1993habitable,kopparapu2013habitable}. Planets within the habitable zone of low-mass stars are expected to fall into synchronous rotation, so that one side of the planet experiences perpetual day with the other side in perpetual night. We use a three-dimensional general circulation model (GCM) to calculate the equilibrium climate state for such synchronously rotating habitable planets at Earth-like and elevated abundances of CFCs. We then \revision{use the steady-state output from these GCM simulations to} calculate synthetic infrared spectra to show that absorption features of these CFCs could be detectable with missions such as JWST and Origins.

\section{Climate Modeling} \label{sec:climate}

The climate simulations in this study are conducted with the ROCKE-3D (Resolving Orbital and Climate Keys of Earth and Extraterrestrial Environments with Dynamics) model \citep{way2017resolving}. ROCKE-3D is a GCM that has been developed by the NASA Goddard Institute for Space Studies for the study of planetary habitability. ROCKE-3D has been used to understand the climate of ancient Venus \citep{way2016venus,way2020venusian}, explore possible habitable climates for specific exoplanets \citep{kane2018climate,del2019habitable,fauchez2019TRAPPIST}, and constrain the dependencies of general habitability limits on planetary properties \citep{way2017effects,fujii2017nir,checlair2019no,colose2019enhanced,salazar2020effect,olson2020oceanographic}. Our configuration of ROCKE-3D assumes aquaplanet (i.e., ocean-covered) conditions with a 1\,bar atmosphere composed of N$_{2}$ and H$_{2}$O with 400\,ppm CO$_{2}$. We used a ``slab'' ocean with a fixed 50\,m depth and a \secrev{present-day Earth} q-flux parameterization of oceanic energy transport. The use of a slab ocean significantly reduces computational time, although the use of a dynamic ocean can cause differences in surface temperature by a few degrees \citep{colose2020effects}. The model includes 40 vertical layers and a $4^{\circ}\times5^{\circ}$ horizontal resolution. We assume a planet with zero obliquity that has the mass and gravitational acceleration of Earth. A more thorough discussion of the model configuration, including details of the radiative transfer methods are provided by \citet{colose2020effects}, while a broader technical description of ROCKE-3D is given by \citet{way2017resolving}.

\begin{figure}
    \gridline{\fig{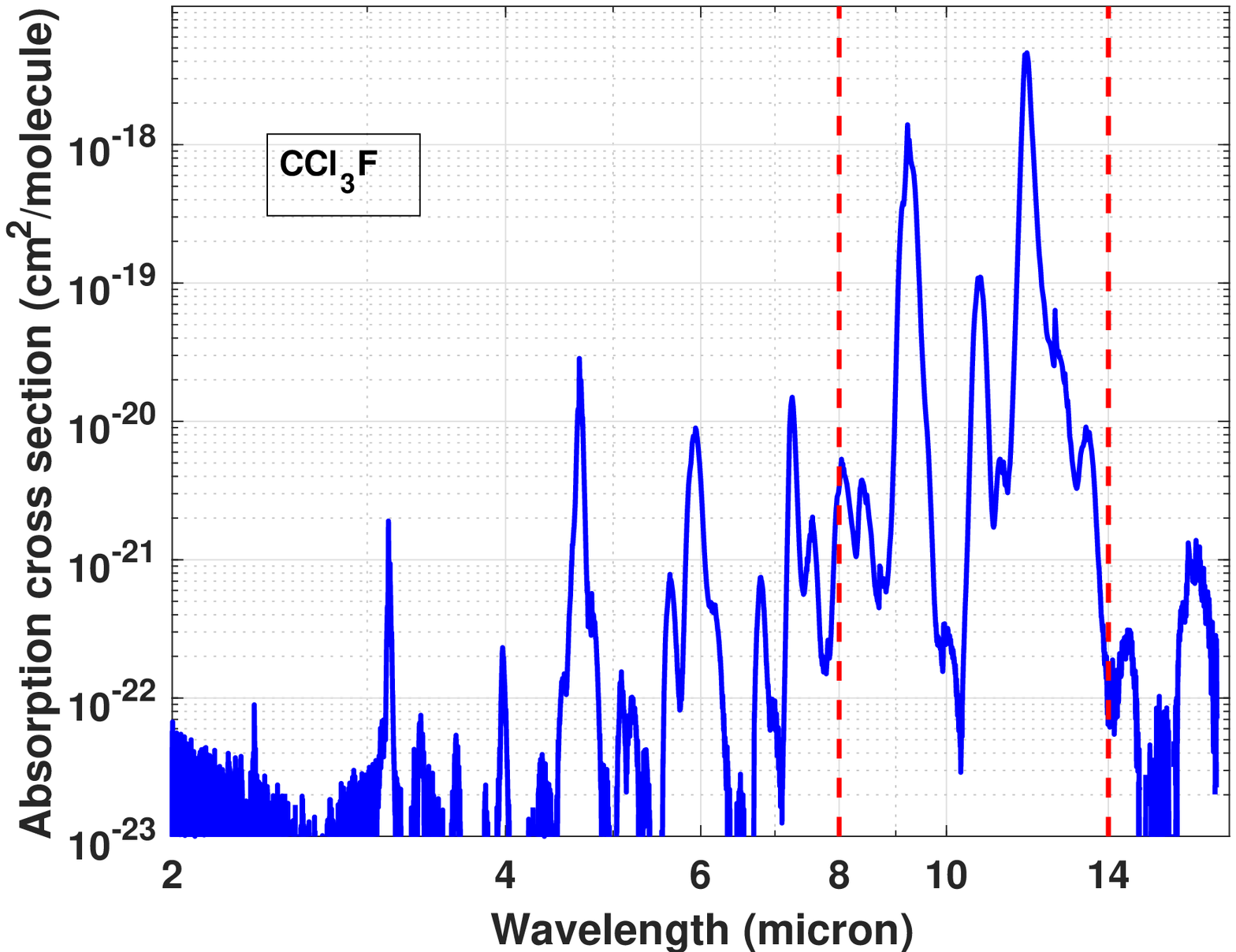}{0.49\textwidth}{}
          \fig{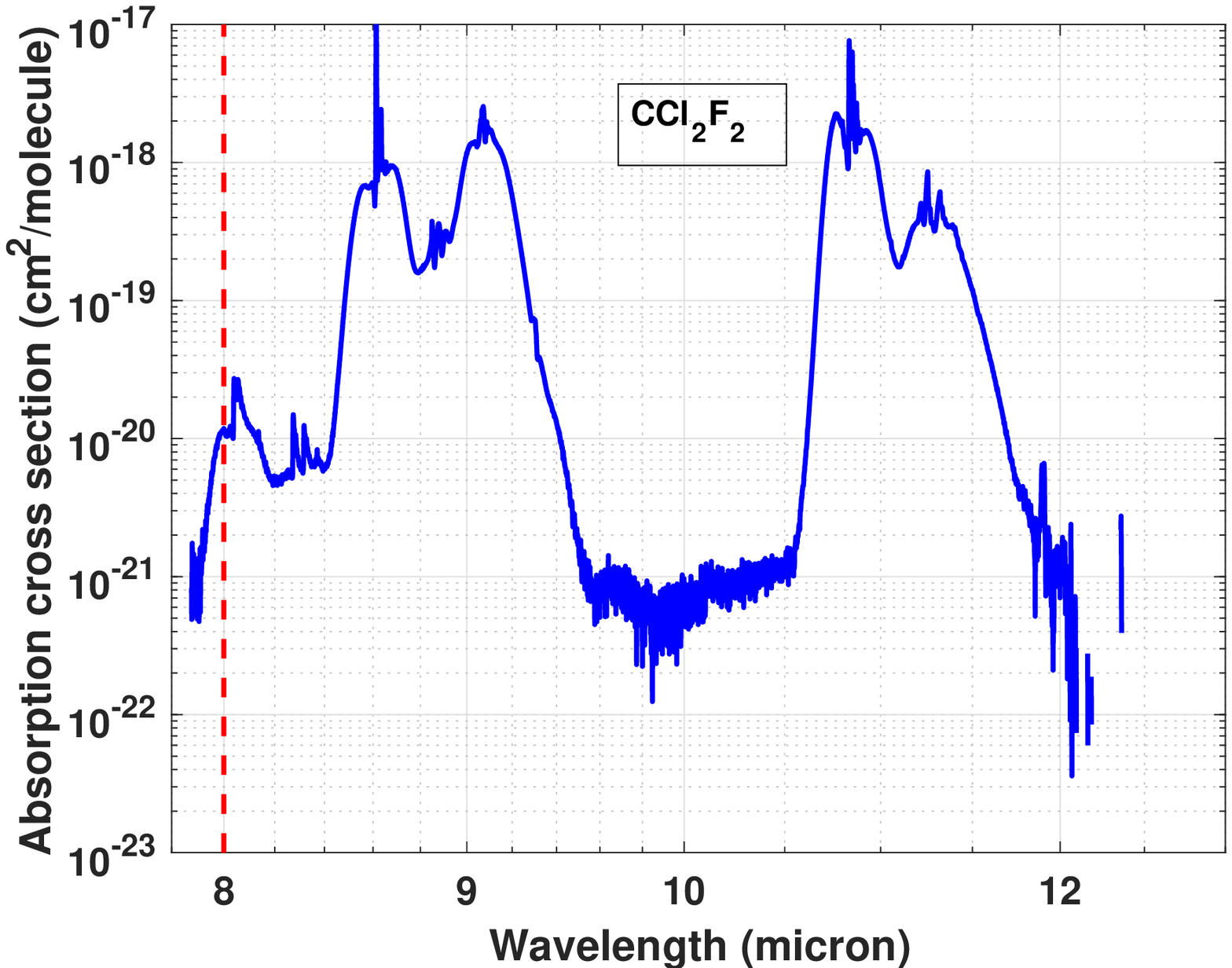}{0.49\textwidth}{}
          }
    \caption{The absorption cross sections for CFC-11 (left) and CFC-12 (right) show \revision{the strongest} features within \revised{a} 8-14\,$\mu$m window \revision{(red dashed lines) \secrev{with one \newrev{major} peak between 8-10\,$\mu$m and a second between 10-14\,$\mu$m}. Cross sections are from the HITRAN database \secrev{with 0.003\,$\mu$m resolution at 300\,K} \citep{sharpe2004gas,harrison2015new,gordon2021hitran2020}.}} 
    \label{fig:cfc}
\end{figure}

\begin{figure}
    \gridline{\fig{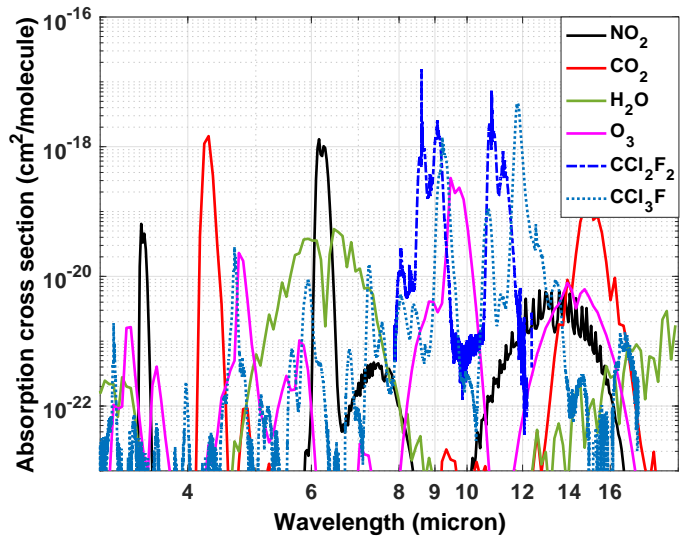}{0.55\textwidth}{}
          }
    \caption{\revision{Examples of} other atmospheric gases \revision{on Earth that} have overlapping absorption features with CFC-11 and CFC-12 within the 8-14\,$\mu$m window. \revision{Cross sections are from the HITRAN database \secrev{with 0.003\,$\mu$m resolution at 300\,K} \citep{sharpe2004gas,harrison2015new,gordon2021hitran2020,gordon2022hitran2020}.}} 
    \label{fig:cfcall}
\end{figure}

Here we perform ROCKE-3D simulations of an Earth-sized planet in synchronous rotation around a 3300\,K and TRAPPIST-1 host stars with increased abundances of atmospheric CFC-11 and CFC-12. These species are included in ROCKE-3D as part of the Earth system model. They have strong absorption bands in the mid-infrared part of the spectrum (Fig. \ref{fig:cfc}); however, this part of the IR spectrum is dominated by several other greenhouse gases such CO$_{2}$, H$_{2}$O and CH$_{4}$ (Fig. \ref{fig:cfcall}) that could make a detection of CFCs difficult with upcoming observatories. In section 3, we will discuss the detectability of CFC-11 and CFC-12 in detail. \secrev{We focus on these two particular systems because they are characteristic of systems likely to actually be studied by missions like JWST and Origins in which CFCs could actually be detectable.}

\begin{figure}[ht!]
\centering
\includegraphics[width=1.00\linewidth]{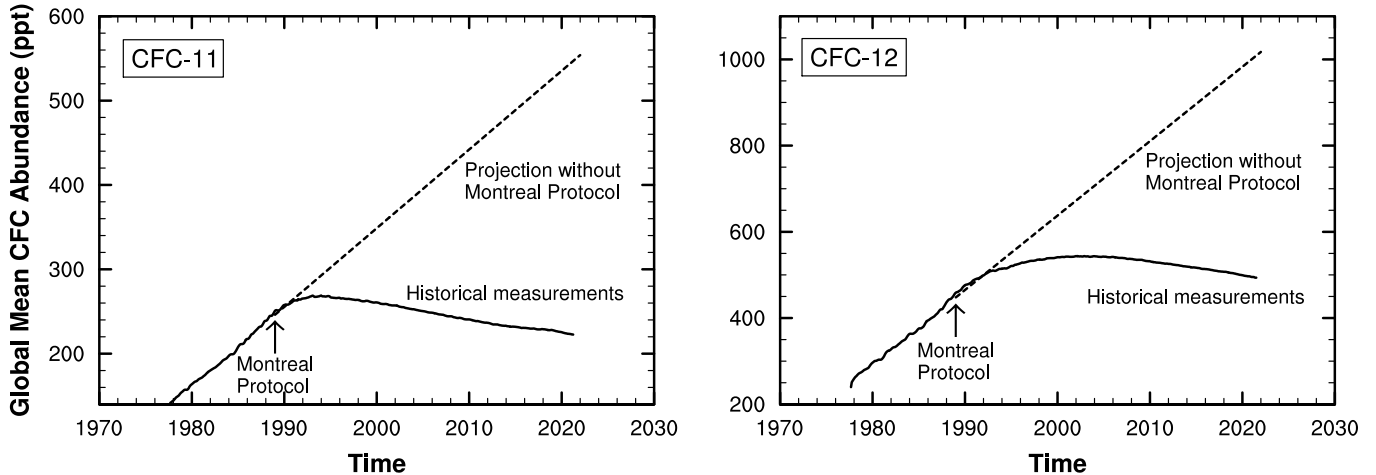}
\caption{Historical measurements of CFC-11 (left) and CFC-12 (right) show a sharp decline that correlates with the adoption of the Montreal Protocol (solid curves). A linear projection shows that CFC abundances of 2 times or greater than today could have been possible if this treaty had not been effective (dashed lines). CFC data is from the National Oceanic and Atmospheric Administration (NOAA) Global Monitoring Laboratory.\label{fig:cfchistory}}
\end{figure}

Our model experiments consist of four simulations conducted with different atmospheric abundances of CFC-11 and CFC-12. \revision{Both of these are potent greenhouse gases with an average lifetime of $\sim$55 years for CFC-11 and $\sim$140 years for CFC-12 in the Earth-Sun system \citep{seinfeld2008atmospheric}}. The 0x case contains no CFCs and provides a reference control case for comparison. The 1x case includes 0.225\,ppb CFC-11 and 0.515\,ppb CFC-12, which are the present-day abundances of these CFCs on Earth.\footnote{This data is from the National Oceanic and Atmospheric Administration (NOAA) Global Monitoring Laboratory, \url{https://www.esrl.noaa.gov/gmd/hats/}.} The subsequent \revision{2x and 5x} cases contain CFC-11 and CFC-12 abundances that have been increased by \revision{2 and 5} times, respectively. \revision{These elevated CFC abundances are realistic projections\secrev{, shown in Fig. \ref{fig:cfchistory},} that could have occurred on Earth if the Montreal Protocol had been ineffective \secrev{\citep[c.f.][]{young2021montreal}}}.

The CFC abundance is fixed in the model, with no sinks or sources due to chemistry. On Earth, the sources of CFC-11 and CFC-12 are industrial sites at the surface, while the only sinks occur when the CFCs rise into the stratosphere and are photolyzed at wavelengths between 185-210\,nm. This results in the relatively long atmospheric lifetimes of CFC-11 and CFC-12 on Earth. Planets orbiting M-dwarfs tend to receive even less incident shortwave radiation on average, so the lifetime of CFCs may be even longer. This study presents a set of \revision{steady-state} calculations with fixed CFC abundances that are intended to constrain the order of magnitude of detectability, but further studies with a coupled chemistry-climate GCM would provide more robust constraints on plausible upper limits for the abundance and lifetime of CFCs.

\revision{We consider two planet-star system configurations, both of which are motivated by previous studies of planetary habitability using GCMs. The first configuration uses a climate configuration for TRAPIST-1e that was defined and studied in previous model intercomparisions \secrev{\citep{fauchez2020trappist,fauchez2021trappist}}. The second configuration uses a climate configuration for a planet in the habitable zone of a 3300\,K host star that has been examined in previous studies \secrev{\citep{kumar2017habitable,colose2020effects}}}.

\subsection{TRAPPIST-1 Host Star}

The seven-planet TRAPPIST-1 system \citep{gillon2017seven} is a target of particular interest by JWST, ELTs, and other future missions, with TRAPPIST-1e orbiting within the star's liquid water habitable zone. Climate modeling studies of the TRAPPIST-1 planets seek to explore possible habitable, and uninhabitable, spectral signatures that could be identified in future attempts at spectral atmospheric characterization \citep[e.g.][]{wolf2017assessing,turbet2018modeling,fauchez2019impact,sergeev2020atmospheric,kane2021phase,may2021water}. Recently, the TRAPPIST-1 Habitable Atmosphere Intercomparison (THAI) workshop established a framework for comparing the capabilities of models to simulate the climate of TRAPPIST-1e, with four general circulation models (including ROCKE-3D and ExoCAM) used as the basis for this intercomparison \citep{fauchez2020trappist,fauchez2021trappist}. This process focused on four specific scenarios for the atmosphere of \revision{TRAPPIST-1e}, which showed general agreement but also revealed some differences in the mean and time-varying states of each of the models. The use of modeling protocols and intercomparisons such as THAI provide a systematic approach toward the use of climate models for understanding exoplanet habitability.

\revision{For the first system in} this study, we use the same ROCKE-3D configuration that was used for THAI  by \citet{fauchez2020trappist} to investigate the detectability of CFCs on TRAPPIST-1e. The model setup assumes a 2590\,K BT-Settl spectrum for the TRAPPIST-1 host star with with luminosity $0.000553L_{\sun}$ and mass $0.0898M_{\sun}$. The planet TRAPPIST-1e is assumed to be in synchronous rotation with a rotational and orbital period of 6.1\,days, an incident stellar flux of $0.662F_{\earth}$, a mass of $0.772M_{\earth}$, and a radius of $0.910R_{\earth}$ (c.f., Table 1 by \citet{fauchez2020trappist}). The model is configured according to the THAI \secrev{Hab1} case, which includes a 1\,bar atmosphere composed of N$_{2}$  with 400\,ppm CO$_{2}$ \secrev{and H$_{2}$O as a variable gas}, as with the 3300\,K cases \revision{described next}. These simulations also use a fixed slab ocean, which remains consistent with the THAI simulations. \revision{Our simulations of TRAPPIST-1e began with 300\,K isothermal initial conditions and ran for 9800 model \secrev{orbits} to reach a statistically steady state, \secrev{and our analysis focuses on an average of the final 1000 orbits.} \secrev{(Note that this long integration time is required when using the ROCKE-3D GCM to simulate synchronously rotating planets with short orbital periods.)}}

\subsection{3300\,K Host Star}

A recent study by \citet{colose2020effects} used ROCKE-3D to calculate the limits of the inner edge of the habitable zone for Earth-sized planets in synchronous rotation around low mass stars with effective temperatures ranging from 4500\,K to 2600\,K. The stellar spectra used by \citet{colose2020effects} were the same set of BT-Settl model spectra from a previous study of the inner edge of the habitable zone by \citet{kumar2017habitable} that used the ExoCAM GCM. Likewise, the simulations performed by \citet{colose2020effects} with ROCKE-3D used the same values of planetary rotation rate and orbital distance that were used in the simulations by \citet{kumar2017habitable}, which were calculated for each host star to remain consistent with Kepler's laws. This approach allowed \citet{colose2020effects} to identify systematic differences between the two GCMs when calculating the inner edge of the habitable zone, which included a comparison of both models configured with a fixed-depth slab ocean. Although some differences arose due to each model's parameterization of cloud formation, the two models generally provided similar constraints on the inner edge of the habitable zone.

\revision{For the second system in} this study, we use the same ROCKE-3D configuration that was used by \citet{colose2020effects} to investigate the detectability of CFCs on planets in the habitable zone of a low-mass star. We focus a set of our simulations on a 3300\,K \revision{BT-Settl} host star with luminosity $0.00972L_{\sun}$ and mass $0.249M_{\sun}$. The choice of a 3300\,K host star falls in the middle of the range of stellar effective temperatures considered by \citet{colose2020effects} and \citet{kumar2017habitable}. The dynamical state of the atmosphere of synchronously rotating planets also changes with its rotation rate, and thus with its distance from the star. An analysis of the  \citet{kumar2017habitable} simulations by \citet{haqq2018demarcating} showed that the 4500\,K to 3300\,K cases reside in a ``slow rotation'' regime with strong heating and convection beneath the planet's substellar point and a much colder night side, whereas the 2600\,K and 3000\,K cases fall into ``rapid'' and ``intermediate'' rotation conditions, respectively, with enhanced zonal energy transport that leads to a warmer planetary night side. The slow rotation regime in particular has been widely examined by others \citep[e.g.][]{joshi1997simulations,joshi2003climate,merlis2010atmospheric,carone2014connecting,kumar2016inner,turbet2016habitability,del2019habitable}. The choice of a 3300\,K host star in this study yields a commonly expected atmospheric state for a synchronously rotating terrestrial planet orbiting within in the habitable zone. 

For this 3300\,K host star, we select a single case of an Earth-sized planet in synchronous rotation with a rotational and orbital period of 22.12\,days and an incident stellar flux of $1.029F_{\earth}$ (c.f., Table 1 by \citet{kumar2017habitable}). This scenario places the planet near the inner edge of the conventional liquid water habitable zone of its host star \citep{kasting1993habitable,kopparapu2013habitable}, but at a distance where the planet retains \revision{habitable surface} conditions and does not show signs of a moist or runaway greenhouse \citep{kumar2017habitable,colose2020effects}. This choice of orbital position is consistent with the assumption in our model that water vapor is the primary atmospheric condensible; by contrast, a terrestrial planet farther toward the outer edge of the habitable zone would reside in conditions where carbon dioxide condensation also occurs. This particular \revision{model configuration} also shows a minimal difference between the use of a fixed slab ocean compared to a dynamic ocean, \revision{as shown in the comparison by \citet{colose2020effects}.} \revision{Our simulations of this 3300\,K scenario began with 300\,K isothermal initial conditions and ran for \secrev{410} model \secrev{orbits} to reach a statistically steady state, \secrev{and our analysis focuses on an average of the final 100 orbits.} \secrev{(Note that synchronously rotating systems in the slow rotation regime require a much shorter integration time using ROCKE-3D compared to planets in a rapid rotation regime.)}}

\section{Synthetic Spectra} \label{sec:OST}

Our GCM simulations \newrev{give} steady-state solutions for atmospheres with fixed abundances of CFC-11 and CFC-12, which can be used to identify the strength of potentially observable spectral features. We use the \revision{calculated outgoing infrared flux values, cloud opacity, and greenhouse gas} mixing ratio values from our GCM simulations as input to the Planetary Spectrum Generator (PSG, \citet{villanueva2018planetary}) to calculate synthetic spectra and assess limits of detectability. PSG is an online radiative transfer software that can be used to compute planetary spectra (atmospheres and surfaces) for various objects of the solar system and beyond. It includes a wide range of wavelengths (UV, visible, near-IR, IR, far-IR, THz, submillimeter, and radio) from any observatory, orbiter, or lander and also includes a noise calculator. In this work, we use a PSG add-on called Global Exoplanet Spectra (GlobES) which allow the user to ingest data from a variety of GCMs to accurately synthesize planetary spectra that are then computed with PSG. GlobES has been used for climate studies of TOI-700d---the first habitable zone terrestrial size planet discovered with TESS \citep{Suissa2020}.

We consider the \revision{GCM \newrev{calculations}} in Section \ref{sec:climate} for planets around TRAPPIST-1 and a 3300\,K host star. \revision{For our synthetic spectra calculations, we use updated parameters for TRAPPIST-1 and TRAPPIST-1e from \citet{Agol2021}. (Although our GCM simulations use older parameters from \citep{Grimm2018}, this discrepancy does not have any quantifiable impact on our results.)} The reason to focus on M-dwarfs is because near-term atmospheric characterization telescope missions focus on planets around M-dwarfs, and their instruments will operate in the IR part of the electromagnetic spectrum. Likewise, as mentioned in earlier sections, CFC-11 and CFC-12 have dominant absorption in the IR, so the confluence of instruments operating in the IR region and the corresponding strong absorption of CFCs makes M-dwarf planets ideally suited for studying CFC detectability. Here we focus on generating the transit spectra of planets around M-dwarfs, and we then estimate the observing time needed to detect the dominant CFC features with JWST and Origins. 

\begin{figure}
    \gridline{\fig{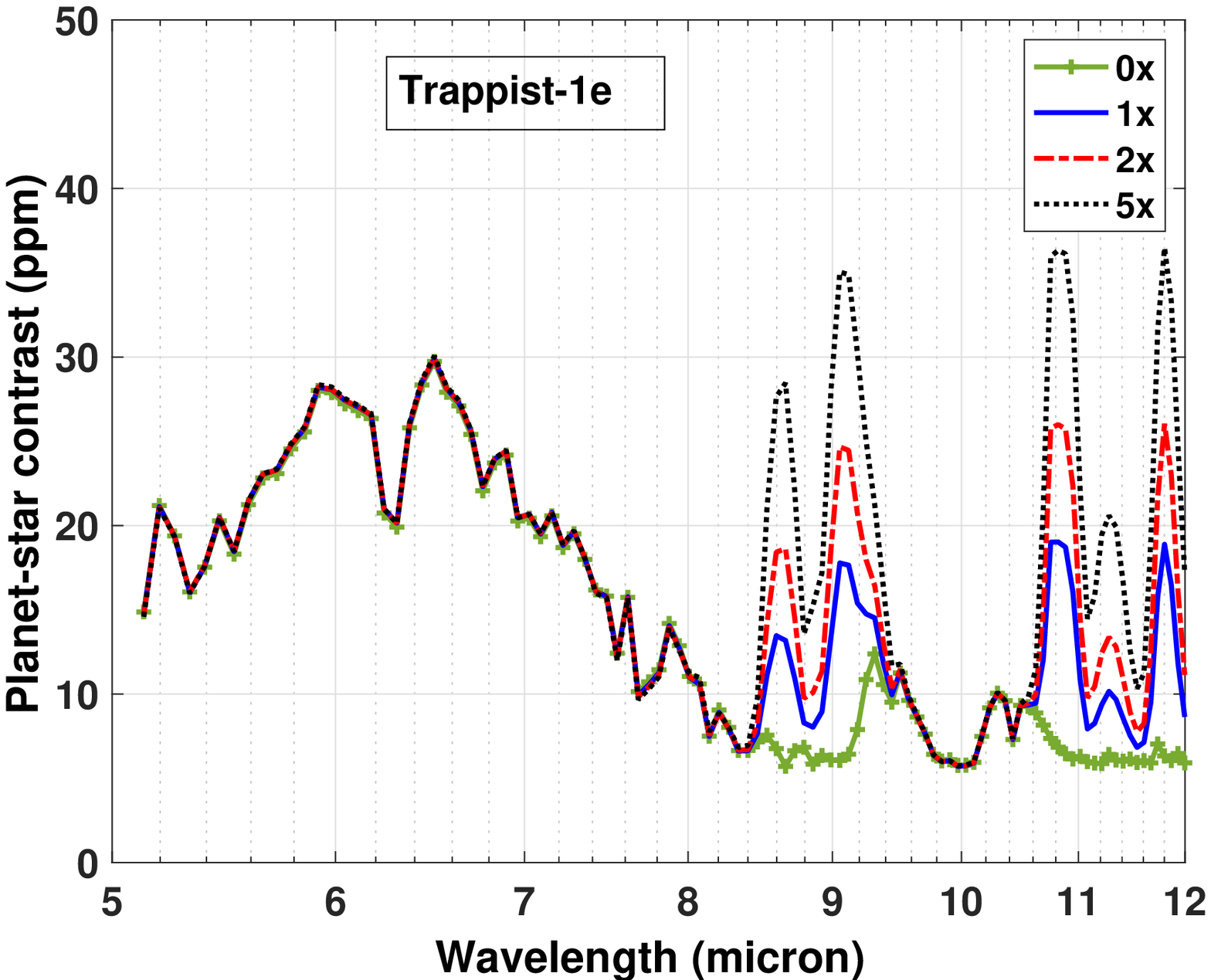}{0.49\textwidth}{a}
    \label{fig:transita}
          \fig{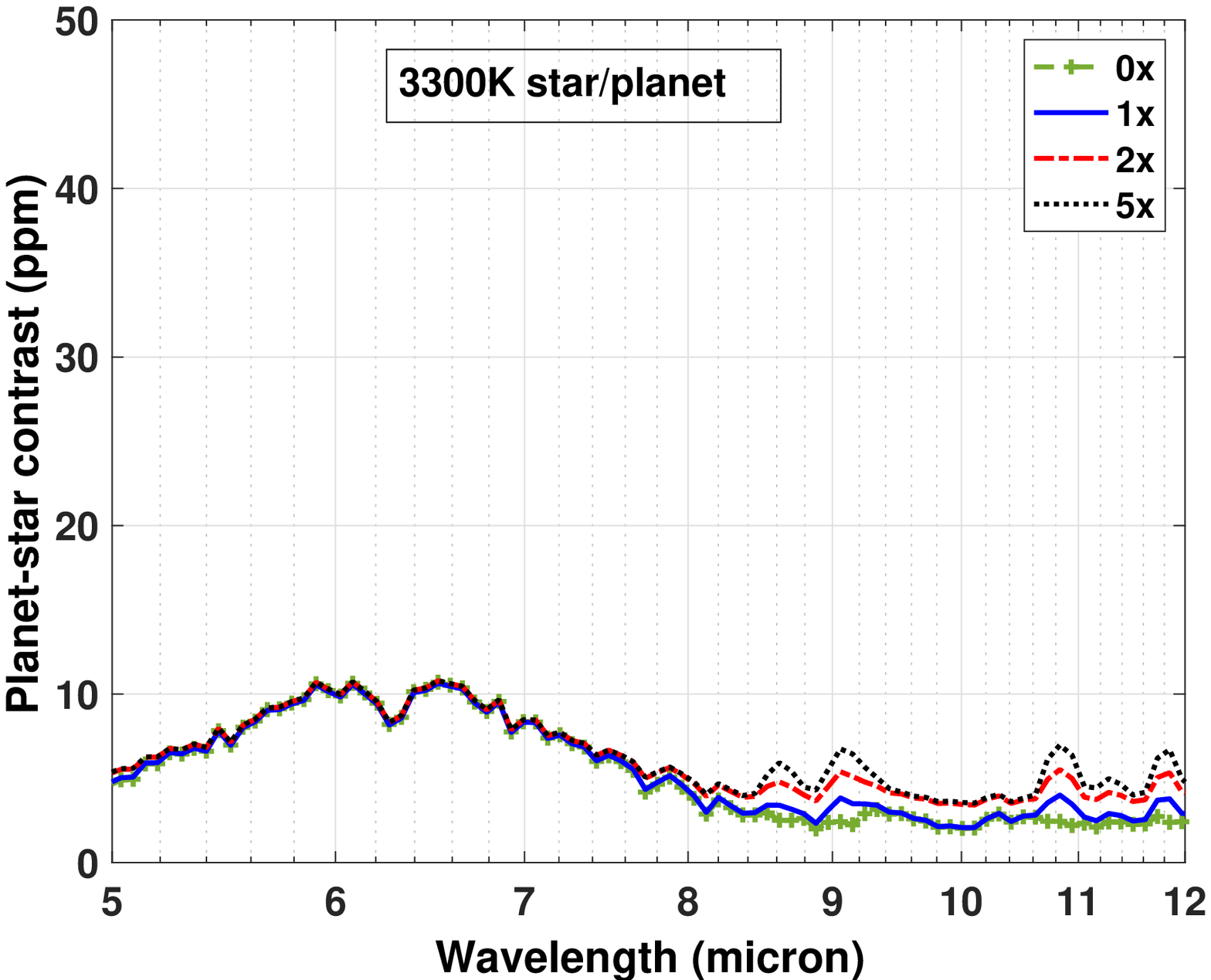}{0.49\textwidth}{b}
          \label{fig:transitb}
          }
    \caption{Transit spectra for TRAPPIST-1e (left) and an Earth-like planet around a 3300\,K host star (right) showing the spectral features of CFC-11 and CFC-12 for the \secrev{0x,} 1x, \revision{2x, and 5x} scenarios. 
    The features are more pronounced around Trappist-1 because of the smaller size of the star. The planets are in the respective habitable zones of their host stars, and the 3300K system is kept at a distance of 5\,pc from the Solar system (TRAPPIST-1 is at a distance of 12.4\,pc).} 
    \label{fig:transit}
\end{figure}

Fig. \ref{fig:transit} shows transit spectra for TRAPPIST-1e and an Earth-like planet around 3300\,K star, both with \secrev{0x,} 1x, \revision{2x, and 5x} Earth-level abundances of CFC-11 and CFC-12. The strongest absorption features are between the \revision{$8-12\,\mu$\,m} range\secrev{.} 
We will focus our detectability estimates within this region. As expected, higher amounts of CFCs increase the transit depth represented as planet-to-star contrast ratio in ppm on the y-axis. The depth of the $8-14\,\mu$m features for TRAPPIST-1e are larger compared to a planet around a 3300\,K star because the stellar host is correspondingly smaller (TRAPPIST-1 is just slightly larger than Jupiter), which would enable relatively more surface area of the star to be transited by the planet and therefore increase the transit depth. This suggests that CFC absorption features might be easier to detect around late M-dwarfs than the earlier ones.

To estimate the signal-to-noise ratio (SNR) for detecting these CFC features, we kept the TRAPPIST-1e planet at a star-planet separation of $0.0293$\,AU and at a distance of 12.4\,pc. We used the TRAPPIST-1e planet parameters from \cite{Grimm2018} \revision{for our GCM simulations as done in  THAI \citep{fauchez2020trappist}.} For the planet around a 3300\,K star, we kept the planet at 0.1\,AU, corresponding to the habitable zone for this star, and at a distance of 5\,pc from Earth. Any SNR calculations can be qualitatively scaled to other distances based on our results for this star type. We have calculated the SNR by subtracting the \revision{0x case with no CFCs
from the 1x, 2x, and 5x} cases 
and dividing the result with the noise, all within a  wavelength range \revision{of the instrument under consideration}. This method evaluates the SNR with respect to the \revision{0x} level of the observed spectra. 

\begin{figure}
    \gridline{\fig{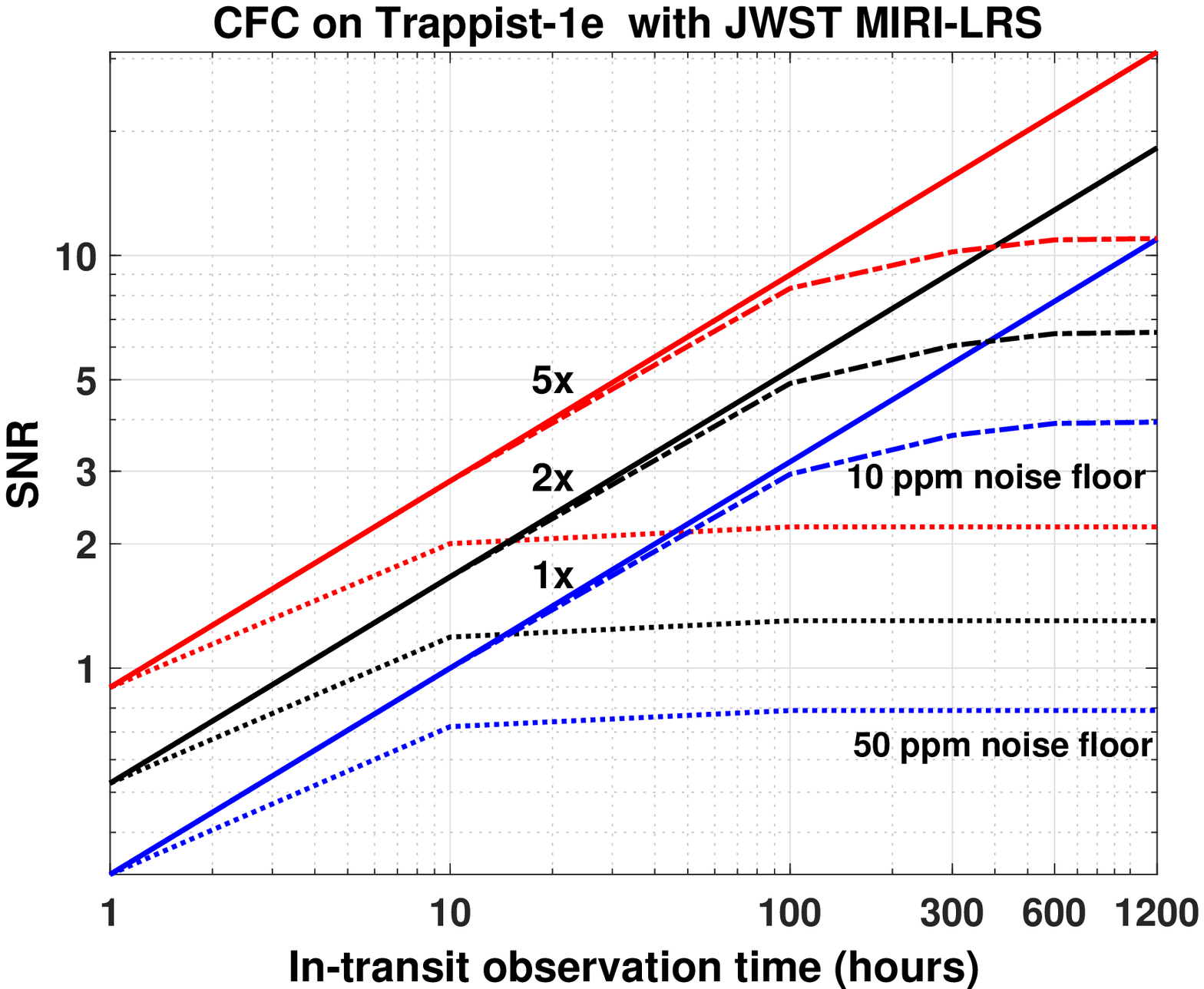}{0.49\textwidth}{(a)}
          \fig{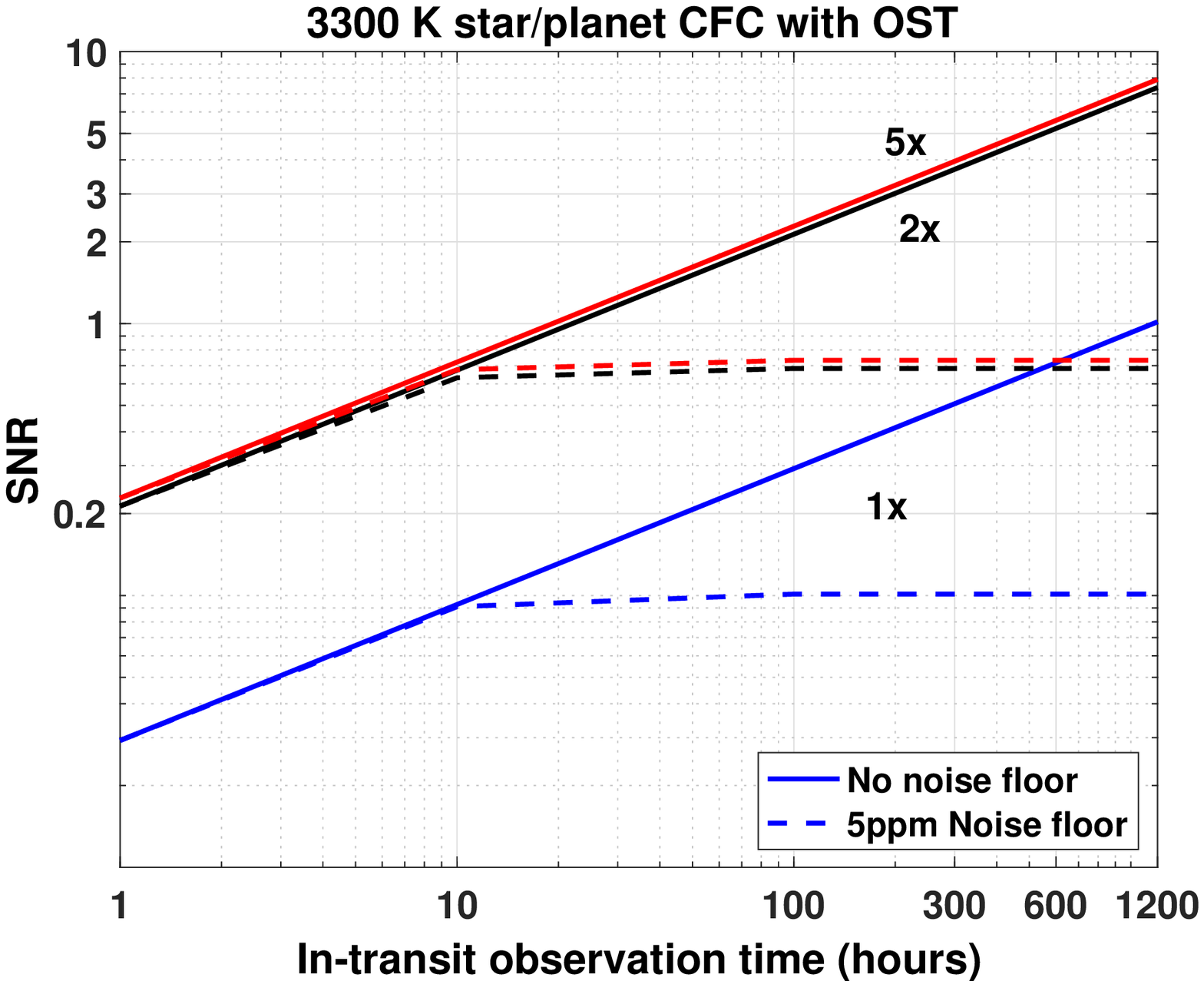}{0.49\textwidth}{(b)}
          }
    \caption{In-transit time (i.e, no overheads) signal to noise ratio estimates as a function of CFC concentration for \revision{TRAPPIST-1e (left) and a planet around a 3300K star (right). The left panel shows JWST MIRI-LRS simulated} observations of TRAPPIST-1e at observing times ranging from 1\,hr to 1200\,hr \revision{with a lower limit of noise floor  of 10\,ppm and higher limit of 50\,ppm. With an optimistic 10\,ppm noise floor, current Earth-level CFCs (1x) could be detectable with a SNR $\sim 3$ in about $\sim 100$ hours of JWST MIRI-LRS in-transit observations. Past Earth-level (2x) CFCs could be detectable on TRAPPIST-1e with a SNR $\sim 5$ for the same amount of time. However, a very conservative 50\,ppm noise floor would render even a 5x present Earth-level CFC abundance challenging to detect. The right panel shows simulated observations with Origins for a planet around a 3300\,K star, with a noise floor of 5\,ppm. Despite the lower noise floor, detection of CFCs around this star would be difficult due to the larger size of the star.}} 
    \label{fig:JWST}
\end{figure}

Fig. \ref{fig:JWST} (left) shows the SNR values for \revision{1x (blue), 2x (black) and 5x (red) CFC versus in-transit observation time in} hours (i.e, no overheads) with JWST MIRI-LRS \revision{(Mid Infrared Instrument low resolution spectrometer)} using a resolution R = 50. \revision{Also shown are two levels of noise floors: 10\,ppm (dashed) and 50\,ppm (dotted). We note that these noise floors are applied to multiple co-added observations, and as such the 50\,ppm floor is expected to be very conservative. With a 10\,ppm noise floor,} concentrations at the current Earth level \revision{(1x, blue-dashed)} can be detectable with SNR $\sim 3$ in $\sim 100$ hours. \revision{While this may not be a robust detection, this could hint at a possible feature for further observations.} However, \revision{the features of a 2x CFC level, which might have been Earth's current CFC levels without the Montreal protocol, would be detectable with a SNR $\sim 5$ with the same amount of time. We would be able to detect the features of a planet with a 5x CFC level (red-dashed) with a SNR $\sim 10$ in 100 hours of JWST MIRI-LRS time on TRAPPIST-1e. However, if the noise floor is set at 50\,ppm, then even the 5x CFC features would not be detectable with a reliable SNR no matter the JWST MIRI-LRS observation time. While upper limits can be placed, these limits will not be robust to make any meaningful analysis.} 


Fig. \ref{fig:JWST} (right)  shows similar SNR calculations for CFC on an Earth-like planet around a 3300~K host star using Origins. \revision{The system is} located at 5\,pc \revision{from Earth, which is intended to provide an optimal scenario for calculating CFC detection limits for Origins. There are about 60 stars within 5\,pc, most of which are M-dwarfs with about half in binaries, all of which are on the Breakthrough Listen target list \citep{isaacson2017breakthrough}.} \revision{The} spectral features  are muted (see Fig. \ref{fig:transit} (right)) owing to a larger star, \revision{and} the SNR values are \revision{lower compared} with TRAPPIST-1e for the present Earth-level case. However, higher CFC concentrations generate much higher SNR values. For \revision{the 2x and 5x} CFC cases, a SNR $\sim 5$ can be achieved with \revision{$\sim 600$ hours} of Origins observing time. 



Note that the estimated integration times only take into account in-transit times. The real observation times including overhead would likely be two or three times larger. Also, no instrument noise systematics were considered in our JWST simulations which only include a pure white noise (i.e. a noise that goes down with 1/$\sqrt{photons}$). The real noise will decrease at a slower rate therefore increasing the required number of transits to achieve 3 or 5 $\sigma$. Also, for long integration times over a large aperture like JWST, instrument systematics are expected to lead to a noise floor, which is a level of noise that can't be reduced by adding more observations \revision{as shown in Fig. \ref{fig:JWST}} \citep{fauchez2019impact}. 

\revision{As mentioned above, around $\sim 100$ hours of JWST MIRI-LRS observing time is needed to detect present Earth level CFCs on TRAPPIST-1e with a noise floor of 10~ppm. It would take observing almost every transit in JWST's mission lifetime to get to $\sim 100$ hours of in-transit time. There are $\sim 100$ transits observable between June 2022-July 2028\secrev{; this} is technically longer than JWST's \secrev{nominal} mission lifetime of 5.5 years\secrev{, but NASA now estimates that JWST ``should have enough propellant to allow support of science operations in orbit for significantly more than a 10-year science lifetime.''\footnote{\url{https://blogs.nasa.gov/webb/2021/12/29/nasa-says-webbs-excess-fuel-likely-to-extend-its-lifetime-expectations/}}} TRAPPIST-1e is likely going to remain the best target for such observations in the foreseeable future, \secrev{so such an investment} would be an expensive but \secrev{potentially} highly-rewarding program.}

Here, we have only calculated whether the effects of CFCs might be detectable at all and what its form would be, and not whether they can be unambiguously retrieved from a real spectrum given realistic ambiguities of the host planet's atmospheric structure and composition, and host star's intrinsic spectrum. We leave an analysis of such complexities to a future study.

In summary, the absorption features of CFC-11 and CFC-12 could \revision{potentially} be detectable by upcoming missions such as JWST, \revision{depending on the noise floor levels. Present or past} Earth-like abundances of CFCs could be detected with observing times of $\sim$100-300\,hr at a SNR \revision{$\gtrsim$ 3-5}. Large observing programs have been conducted previously, such as $\sim$400\,hr for Hubble Ultra Deep Field \citep{beckwith2006hubble} or $\sim$900\,hr for the CANDLES galaxy evolution survey \citep{grogin2011candels}, so this requirement remains plausible. Also, such large observing programs are smaller compared to the  estimates for biosignatures in a modern Earth -like atmosphere on TRAPPIST-1e \citep{fauchez2019impact, Jake2019}, i.e. $\sim$600\,hr for O$_3$ detection at 9.6$\mu m$ or $\sim$800\,hr for O$_2$ at 6.4$\mu m$, with CH$_4$ and H$_2$O being undetectable in this scenario. An interesting point here is that the time needed to detect some \secrev{present-Earth} biosignature gases ($\sim600$ hours) is larger than the observing time needed to detect \secrev{present-Earth abundances of CFCs} ($\sim 300$ hours) with JWST. Furthermore, any attempt at characterizing spectral technosignatures would be conducted in tandem with a more general effort to characterize a planet's atmosphere and identify any potential biosignatures. Calculations such as those presented in this paper are useful in determining observability thresholds for detecting particular technosignatures, such as CFCs, which can aid in the development of observing strategies as well as motivate the design of new technology for future missions.

\section{Discussion} \label{sec:discussion}

Halocarbons such as CFC-11 and CFC-12 are industrial compounds on Earth and thus could be evidence of extraterrestrial technology if observed in an exoplanet atmosphere. In the event of such an observation in exoplanet spectra, any absorption features claimed to be technosignatures would need to be examined against non-technological and non-biological alternatives. On Earth, there are no known abiotic or non-anthropogenic sources of CFC-11 or CFC-12 \secrev{\citep{seinfeld2008atmospheric}, likely} \revision{because such molecules are thermodynamically challenging to produce}; however, this does not necessarily mean that CFC-11 or CFC-12 could not be generated on an exoplanet through non-technological means. For example, other halocarbon species can be emitted to the atmosphere by phytoplankton, although these tend to be short-lived \citep{lim2017halocarbon}. This is an instance of a more general problem in biosignature science, as identifying false positives that occur on exoplanets but not on Earth is a challenging task. Nevertheless, advancing the science of technosignatures will require evaluating such false positives for technosignatures such as CFCs, just as biosignature science continues to evaluate false positive spectral signatures for habitability.

The calculations presented in this study indicate that CFCs could \revision{potentially} be detectable by upcoming missions and future mission concepts \revision{given noise floor constraints}, even at present-day Earth levels. \secrev{The detection of absorption features within the 8-14\,$\mu$m region, and subsequent identification of CFCs as the source of these features, will remain challenging, and future work will be needed to improve constraints on the detectability of CFCs for specific targets as observed by JWST, Origins, or other missions. This study focused on the detectability of two M-dwarf systems that could plausibly be observed by JWST or Origins, but future work could examine the possible detectability of CFCs in the atmospheres of habitable planets around solar-type stars with other mission concepts, such as LIFE.}

\secrev{We note that our study has focused on the concept of ``detectability'' for CFCs, and the actual detection of absorption features within the 8-14\,$\mu$m region would not themselves be sufficient evidence to conclude the presence of CFCs in an exoplanet atmosphere. For an atmospheric composition like present-day Earth, CFCs may show some unique spectral features that help distinguish it from other gases (Fig. \ref{fig:cfcall}), but other molecules such as CH$_4$, NH$_3$, CO, H$_2$S, and PH$_3$ also have absorption features that overlap with those of CFC-11 and CFC-12. Resolving any ambiguity regarding the identity of absorbing species will require further observations at other wavelengths that resolve additional absorption features. Theoretical modeling can also help to provide constraints on possible false positive (and false negative) scenarios for detecting CFCs in the atmospheres of exoplanets.}

One limitation of this modeling study is that CFC abundances were assumed to be fixed and scaled without consideration of atmospheric chemistry. The use of a coupled chemistry-climate model would allow for a more self-consistent prediction of CFC abundances that could be sustained in the atmosphere, which would depend on the atmospheric composition as well as the stellar spectrum. \revision{Our GCM cases also assumed a uniform distribution of CFCs across the atmosphere, which assumes that localized sources \secrev{or sinks} of CFC production are well-mixed over time. However, scenarios in which strong, localized, and continuous sources \secrev{and sinks} of CFCs dominate other sources \secrev{and sinks} on the planet would require further study \secrev{with a coupled climate-chemistry model to more accurately constrain the detectability of CFCs}.}

It is worth asking whether or not it is even reasonable to consider the detectability of planets with CFC abundances much greater than those on Earth today. Governments on Earth have banned the use of CFCs that could deplete ozone, while the threat of exacerbating climate change keeps our civilization from allowing long-lived radiatively active CFCs to accumulate in significant quantities. Such risks could also motivate extraterrestrial civilizations to minimize use of CFCs, although this is strongly dependent on the details of that species’ climate needs, the planet’s atmospheric composition, and many assumptions about the species’ long-term aims and coordination. For instance, the species might be incapable of preventing the buildup (because it is insufficiently organized or motivated to), indifferent to the buildup (because it has no important effects on the planet or because they are unaware of those effects), or causing the buildup deliberately (because it wishes to warm the planet, perhaps).

This study focused on planets in M-dwarf systems because such systems are likely to be characterized by upcoming missions, and the steady-state photochemistry on such planets will differ due to lower reaction rates for many atmospheric constituents. For example, planets orbiting K- and M-dwarf stars can accumulate O$_3$ more easily \citep[e.g.][]{segura2010effect,arney2019k}, which could allow for CFCs to accumulate to higher levels than on Earth before causing environmental problems. Such scenarios should be considered in future work and should also be expanded to include a broad range of halocarbons beyond CFC-11 and CFC-12 only. \revision{Similarly, this study focused on the detectability limits of upcoming space missions, but future high-resolution ground-based spectroscopy facilities, such as Extremely Large Telescopes \citep[e.g.,][]{cavarroc2006fundamental,2013ApJ...764..182S, kuhn2015global, 2017A&A...599A..16L, birkby2018spectroscopic}, may also be able to detect the presence of CFCs in exoplanet atmospheres.}


We do not know the extent to which the specific CFCs produced on Earth would be prevalent elsewhere, even for extraterrestrial civilizations with similar industrial processes. The family of halocarbons is large, and much more work would be needed to assess the detectability of a broader range of industrial molecules. Such an effort would be a step toward constructing a library of technosignatures for planning and interpreting future observations. But even such a library may be unable to identify industrial molecules that are chemically possible but have never been generated on Earth. Efforts to explore other possible industrial molecules, as well as their spectral signatures, could also help to constrain the use of halocarbons in general as technosignatures.

\section{Conclusion} \label{sec:conclusion}

CFCs are a notable example of a technosignature on Earth, and the detection of CFCs on a planet like TRAPPIST-1e would be difficult to explain through any biological or geologic features we know of today. Our civilization continues along a path of growth in both population and energy consumption, while we are only beginning to understand the extent to which our technology could be detectable at astronomical distances. Continued exploration of how the past, present, and future of civilization will affect Earth's detectability remains an important objective for understanding the prevalence of biosignatures and technosignatures in our galaxy.

\secrev{In this study,} we have shown that with the launch of JWST, humanity \revision{may} be very close to an important milestone in SETI: one where we are capable of detecting from nearby stars not just powerful, deliberate, transient, and highly directional transmissions like our own (such as the Arecibo Message), but consistent, passive technosignatures of the same strength as our own. Note that this is not a symmetric situation: the detectability of CFCs in an Earth-like planet's atmosphere is strongly dependent on the radius and spectrum of the host star, and the TRAPPIST-1 system in particular is extremely favorable in that regard. Even if Earth were seen to transit from that system, the Sun's large radius and Earth's \secrev{orbital distance} mean that JWST would not detect CFCs around Earth from the distance of TRAPPIST-1.

In the next few decades there will be at least two of Earth's passive technosignatures, radio emissions and atmospheric pollution, that would be detectable by our own technology around the nearest stars. It is possible that other plausible atmospheric technosignatures will prove even more detectable once their signal strengths have been calculated. We conclude that atmospheric technosignatures are at least as promising as communicative, radio technosignatures in this regard, especially since they can be searched for concurrently with biosignatures.

\acknowledgments
 The authors thank Tom Greene and Knicole Colon for their suggestions regarding JWST instrument and noise floor specifics. J.H.M., A.F., J.T.W., and M.L. gratefully acknowledge support from the NASA Exobiology program under grant 80NSSC20K0622. R.K and T.F acknowledge support from NASA Goddard's Sellers Exoplanet Environments Collaboration (SEEC), which is supported by NASA's Planetary Science Division's Research Program. This work was facilitated through the use of advanced computational, storage, and networking infrastructure provided by the Hyak supercomputer system at the University of Washington. Any opinions, findings, and conclusions or recommendations expressed in this material are those of the authors and do not necessarily reflect the views of their employers or NASA.

\software{The ROCKE-3D source code is freely available at \url{https://simplex.giss.nasa.gov/gcm/ROCKE-3D/}. The Planetary Spectrum Generator is available online at \url{https://psg.gsfc.nasa.gov/}. The NCAR Command Language \citep{brown2012ncar} and CET Perceptually Uniform Colour Maps \citep{kovesi2015good} were used in post-processing.}




\bibliography{main}{}
\bibliographystyle{aasjournal}



\end{document}